\documentclass[showpacs,preprint,amsmath,amssymb,showkeys,unsortedaddress,floatfix]{revtex4}

\usepackage[english]{babel}
\usepackage{graphicx}
\usepackage{dcolumn}
\usepackage{bm}
\usepackage{color}

\begin{document}

\preprint{\today}

\title{Hot electrons induced by (rather) cold multiply charged ions}

\author{Thorsten Peters}
\author{Christian Haake}
 \affiliation{Experimentelle Physik, Universit\"at Duisburg-Essen, 47048 Duisburg, Germany}
\author{Detlef Diesing}
 \affiliation{Physikalische Chemie, Universit\"at Duisburg-Essen, 45117 Essen, Germany}
\author{Domocos Kovacs}
 \affiliation{Experimentalphysik II, Ruhr-Universit\"at Bochum, 44801 Bochum, Germany}
\author{Artur Golczewski}
\author{Gregor Kowarik}
\author{Friedrich Aumayr}
 \affiliation{Institut f\"ur Allgemeine Physik, Technische Universit\"at Wien, A-1040 Vienna, Austria}
\author{Andreas Wucher}
\author{Marika Schleberger}
 \email{marika.schleberger@uni-due.de}
 \affiliation{Experimentelle Physik, Universit\"at Duisburg-Essen, 47048 Duisburg, Germany}

\begin{abstract}
Thin film metal-insulator-metal junctions are used in a novel approach to investigate the dissipation of potential energy of multiply charged ions impinging on a polycrystalline metal surface. The ion--metal interaction leads to excited electrons and holes within the top layer. A substantial fraction of these charge carriers is transported inwards and can be measured as an internal current in the thin film tunnel junction. In Ag--AlO$_{\mathrm{x}}$--Al junctions, yields of typically 0.1--1 electrons per impinging ion are detected in the bottom Al layer. The separate effects of potential and kinetic energy on the tunneling yield are investigated by varying the charges state of the Ar projectile ions from 2+ to 9+ for kinetic energies in the range from 1 to 12~keV. The tunneling yield is found to scale linearly with the potential energy of the projectile. 
\end{abstract}



\maketitle

The interaction between energetic ions and solid surfaces has been studied intensively in recent years. Due to the possible promising technological applications the relevant physical processes pose an ongoing challenge for basic research. Ions may carry energy into the solid either as kinetic energy due to their velocity or as potential energy due to their charge state. The ion is slowed down by elastic (nuclear stopping) and inelastic interactions (electronic stopping) within the material while transferring its energy to the solid. 
Material modifications such as sputtering, melting, amorphization, evaporation, etc. require the movement of atoms, i.e.~excitations of the lattice. This may occur directly by nuclear stopping \cite{duvenbeck05_1} or via the excitation of the electronic system by electronic stopping \cite{akcoeltekin2007}. In general, electronic stopping is dominant for fast ions ($v_{\mathrm{ion}}>v_{\mathrm{Bohr}}$), whereas nuclear stopping is dominant for slow ions ($v_{\mathrm{ion}}\leq v_{\mathrm{Bohr}}$). 

The excitation of target electrons with slow ions can be enhanced if multiply charged ions (MCI) are used \cite{gillaspy01,aumayr07}. The interaction of the MCI with the surface leads to a hollow atom \cite{briand90}
which is formed just in front of the surface. Subsequent Auger transitions lead to the emission of electrons partially into the vacuum~\cite{das93}. Many experiments made use of this external emission to study the deexcitation of MCI because ejected electrons can be measured by common methods of electron spectroscopy (for an extensive review see \cite{arnau1997}). However, most of these electrons are already emitted before the projectile has entered the solid. Despite this electron emission the projectile still carries a significant amount of its energy into the solid. Just recently, it has been shown that a MCI deposits almost 90\% of its potential energy within the solid \cite{kost07}. 

So far, a direct investigation of the electronic excitations induced by the potential energy of the projectile within the solid has not been possible. 
The subsurface processes can in principle as well be studied by measuring the external emission (see e.g.~\cite{meyer1991}) but with the disadvantage that only electrons having sufficient energy to overcome the work function of the metal (4.26 eV for polycrystalline silver~\cite{michaelson77}) can be detected. In this paper, we will show that by means of a thin film metal-insulator-metal (MIM) junction, complementary and hitherto inaccessible data can be obtained. With this new approach it is for the first time possible to study the deexcitation of a hollow atom directly {\em within} the solid.

The MIM junctions used here are produced {\it ex situ} by evaporating a 50~nm thick aluminum electrode onto an insulating glass substrate. In an electrochemical treatment described in detail elsewhere \cite{diesing2003}, the aluminum is locally oxidized to form an alumina overlayer of 3.5~$\pm $ 0.5~nm thickness. In a last step, a polycrystalline silver layer of about 25~nm thickness is deposited on top of the oxide layer. Both, its thickness and surface roughness ($rms$ typically $\approx$ 3~nm), are determined by Atomic Force Microscopy (AFM) line scans (not shown here). 

The silver layer has to be thick enough to avoid any penetration of the projectile into the insulating oxide layer. This would change the tunnel barrier in the course of the experiment, thus leading to unreproducible results. For kinetic energies $E_{\mathrm{kin}}\leq 12$~keV, as applied in our experiments, penetration depths of Ar ions of less than (8.5 $\pm $ 5)~nm are predicted~\cite{srim}. On the other hand, the layer must be thin enough to permit a quasi-ballistic transmission of hot charge carriers (electrons and holes) generated at the surface to the oxide interface, where they can tunnel across the junction and be detected as an ion impact induced tunneling current in the underlying Al substrate.

Another important feature of the MIM is the homogeneity of the layer thickness. The minimum thickness of each layer defines its performance. To check the integrity of the MIM we measured the resistance (typically 9 $\Omega$ -- 12 $\Omega$ for the top layer and 6 $\Omega$ for the aluminum, the oxide layer's differential resistance of $>$ 1 G$\Omega$ at 0~V bias) and the capacity (typically 100~nF) of the MIM before and after irradiation.

The samples were irradiated with Ar$^{\mathrm{q+}}$ ions produced by an Electron Cyclotron Resonance (ECR) ion source \cite{ECRIS}. The ion beam was pulsed with typical pulse widths between 150~ms and 1.5~s. By adapting the extraction potential of the ECR-source the samples were irradiated with ions at a constant kinetic energy despite the varying charge state. The charge state was selected by a bending sector magnet. All irradiations took place under UHV conditions (typically $p=5 \times 10^{-8} $ mbar). The angle of incidence was 90$^{\circ }$ with respect to the surface. 

The primary ion current of about 1~nA corresponds to a maximum fluence of $10^{10}$ particles per second with a typical beam cross section of $\pi \cdot 1$~mm$^2$ yielding 10$^{-3}$ ions per surface atom and second. This value is sufficiently low to ensure that ion impact events are spatially and temporarily separated. Therefore, we can treat all data as an average over a great number of electronic excitation events each caused by the impact of an individual ion. 

During irradiation, the surface is subject to sputtering. Measured sputter yields range from about 3 to 10 sputtered Ag atoms per incoming Ar ion with kinetic energies between 5~keV and 12~keV, respectively \cite{andersen}. In the case of metals, the sputter yield does not vary much as a function of projectile charge state as has been shown recently~\cite{aumayr04}. A monolayer of silver contains about $10^{13}$ atoms in the irradiated area. Therefore, with a beam current of 1~nA, less than 0.5\% of a monolayer per second of irradiation time is removed, and more than 100 ion pulses of maximum duration can be measured before one monolayer of silver is sputtered away.  This is in good agreement with our AFM measurements before and after irradiation which showed no significant topological change of the top electrode. The surface roughness remained 3~nm, corresponding to a relative roughness of about 10\% for the top layer. 

\begin{figure}
\includegraphics[width=5cm]{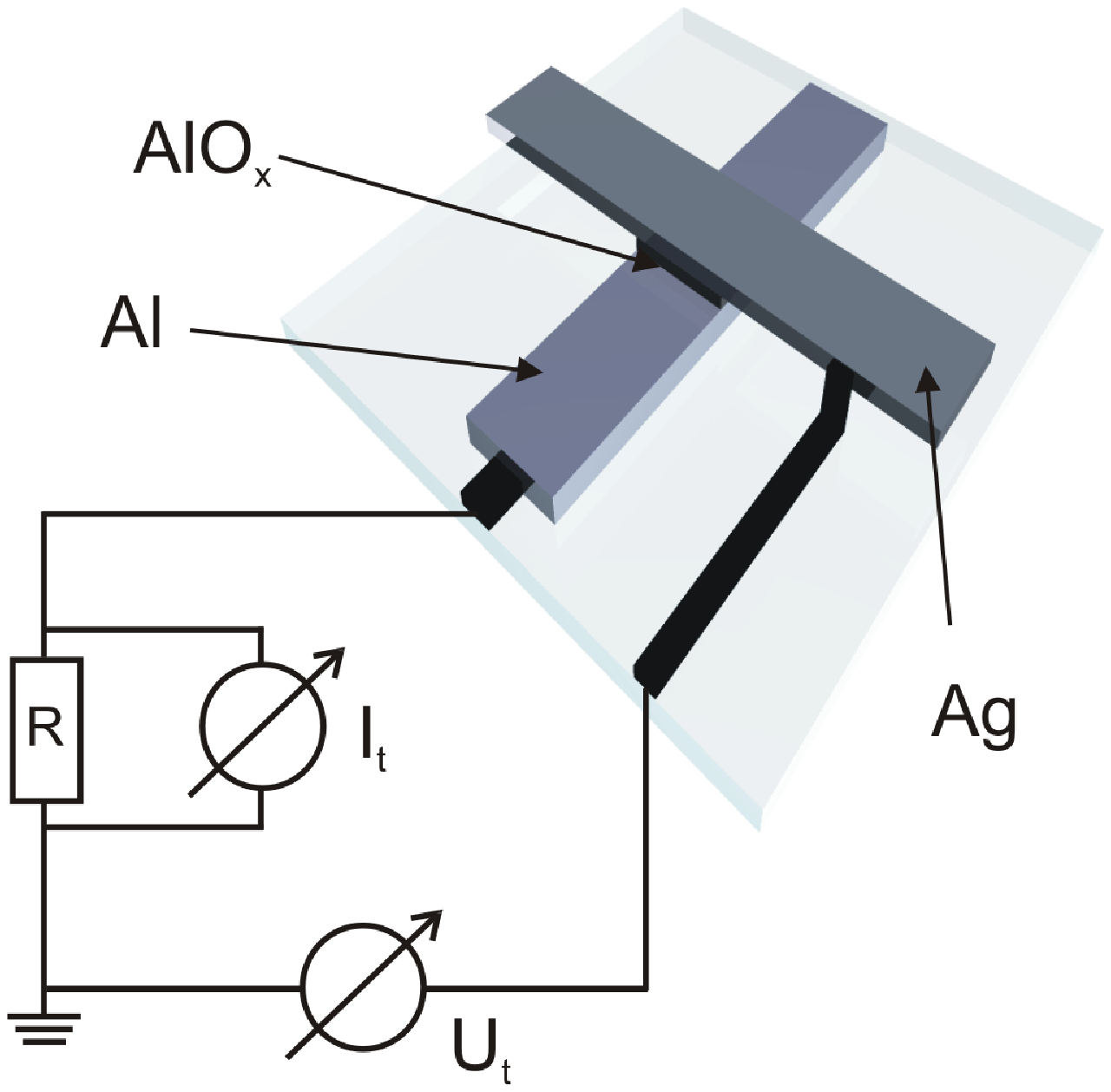}
\includegraphics[width=10cm]{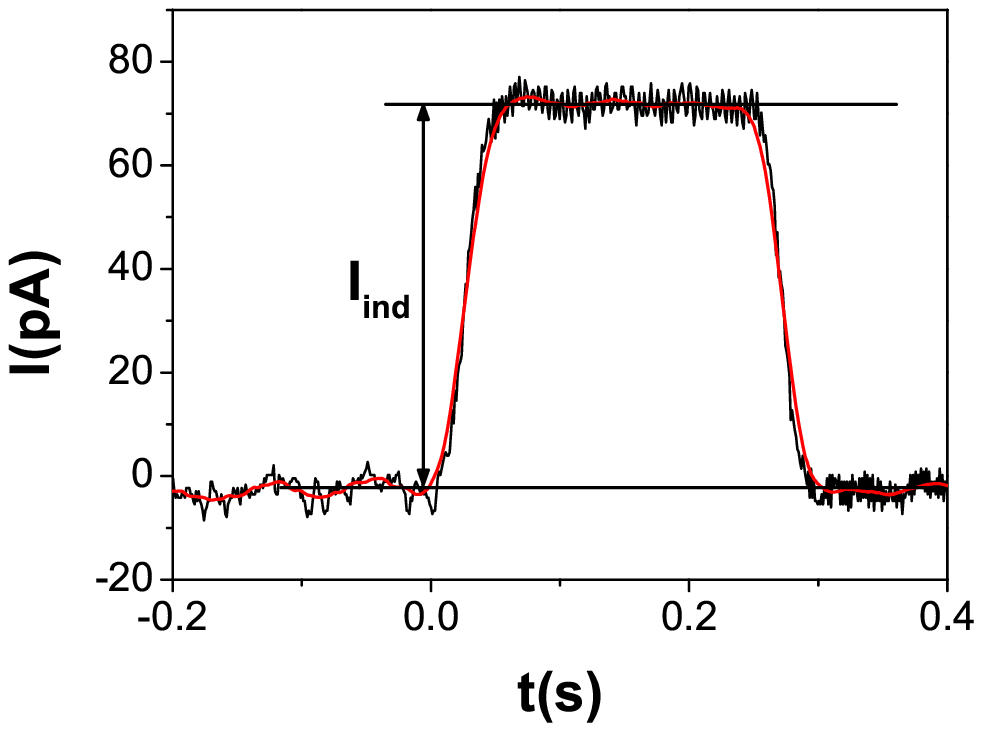}
\caption{Left: Schematic drawing of a MIM junction. Right: $I(t)$-plot for 4~keV Ar$^{4+}$ recorded during one single
irradiation shot with $t_{\mathrm {pulse}} \approx 0.3$~s. In this example the measured current of 76~pA in the bottom electrode corresponds to a voltage of 760~mV. With a measured impinging current of Ar$^{4+}$ projectiles of 1050~pA a tunneling yield of $\gamma=$0.29 was derived from the above formula.
}  
\label{dI} 
\end{figure}

To determine the tunneling yield $\gamma$, i.e. the number of tunneled electrons per primary ion impact, the ion beam current was measured with a Faraday cup right before and after irradiation. The primary ion beam current was typically between 0.1~nA and 10~nA and was stable within 5\% during the irradiation. 
The tunneling current measurements were performed using a $10^{10}~I$-$V$ current converter, the output of which was displayed and recorded as $V(t)$-plot with a digital oscilloscope. The signal-to-noise ratio was typically $> 10 $. Experiments were repeated three times for charge states up to $q=8$ and twenty times for $q=9$ (due to the low current of 50 pA) in order to get reasonable error bars. Fig.~\ref{dI} shows a typical $I(t)$-plot from a MIM during an ion pulse. From the pulse response $I_{\mathrm {ind}}$ and the beam current, the tunneling yield $\gamma=I_{\mathrm {ind}}/(I_{\mathrm {beam}}/q)$ was extracted. 

\begin{figure}
\includegraphics[width=14cm]{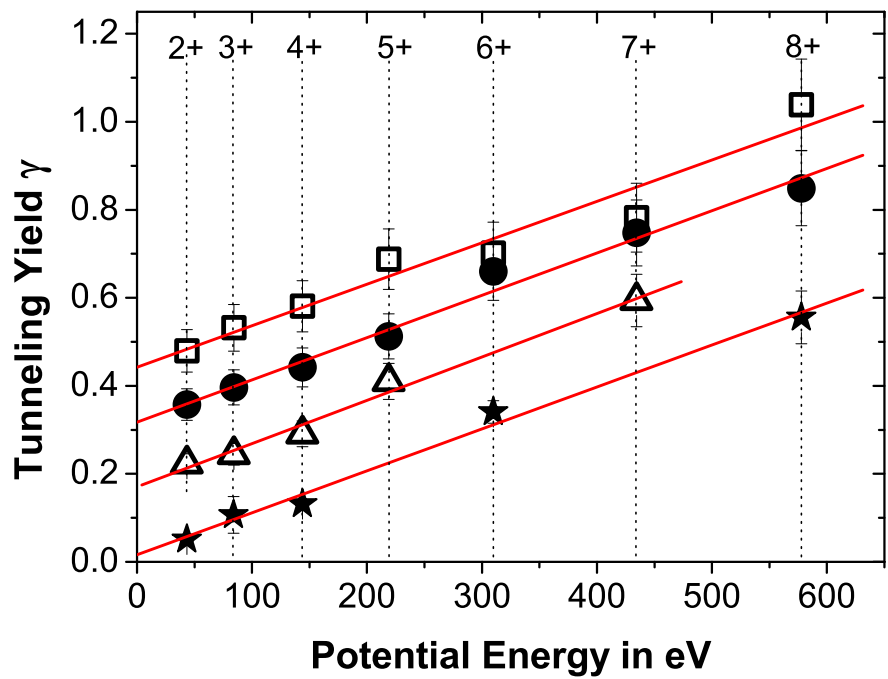}
\caption{Tunneling yield $\gamma $ as a function of potential energy measured for argon projectiles with kinetic energies of 1~keV (stars), 4~keV (triangles), 8~keV (dots), and 12~keV (squares).}
\label{alles} 
\end{figure}

Fig.~\ref{alles} shows the tunneling yield $\gamma $ as a function of potential energy $E_{\mathrm {pot}}$ for four different kinetic impact energies. The data are plotted against the total potential energy stored in the projectile in form of its ionization energy. For each kinetic energy the data are well fitted by a linear function. The slope of each regression line for 1, 4, 8, and 12 keV can be determined within an error of less than 10~\% to $9.5\cdot 10^{-4}$ eV$^{-1}$, $9.7\cdot 10^{-4}$ eV$^{-1}$, $9.6\cdot 10^{-4}$ eV$^{-1}$ and $9.4\cdot 10^{-4}$ eV$^{-1}$, respectivly. 
Consequently, the slope is nearly independent on the kinetic energy of the projectile.


As already mentioned, most of the potential energy is dissipated within the solid \cite{schenkel1999,kost07}. Within the top layer of the MIM junction the hollow atom creates an intense local excitation of the electronic system which we can detect as a tunneling current. Because of the low atomic number of the projectile the deexcitation happens almost exclusively via non-radiative processes \cite{arnau1997,kost07,fink66,wentzel27} 
so we can neglect the amount of energy dissipated by x-ray emission. Thus, the electronic excitation within the top layer of the MIM junction occurs mainly via Auger transitions. 

The energy involved in these Auger processes is carried by several hot charge carriers towards and partially through the oxide. 
In the energy range of 10--200 eV (corresponding to possible Auger transitions), most hot carriers will scatter with other carriers along their way, because the inelastic  mean free path ($\lambda=$ 0.5--3~nm ~\cite{seah79,penn87,tanuma99,ziaja06}) is smaller than the distance between the location of excitation, i.e. the surface and the bottom layer. At the interface the oxide layer acts as a barrier and the transmission probability depends on the energy of the carriers. If the energy is larger than the average barrier height ($\Phi \approx 3$~eV for electrons \cite{diesing2003} and $\Phi \approx 4$~eV for holes \cite{Jennison2004}), the carriers will pass with a probability close to one. For lower energies, the carriers have to tunnel through the barrier. The tunneling probability decreases exponentially with decreasing carrier energy. This latter contribution can be neglected, because the high energy of the excitation events leads to a relatively high population of states above the barrier~\cite{meyer04}.

From fig.~\ref{alles} we find that the number of hot carriers which are able to cross the barrier increases linearly with potential energy. In the range studied, a net number of  $\approx 0.1$ hot electrons per 100~eV potential energy are detected. This increase of the yield is obviously unaffected by the kinetic energy of the projectile. Note, that image charge acceleration cannot be the reason for the observed increase because in our case the corresponding energy gain is well below 30~eV even for the highest charge state \cite{burgdoerfer93,winter92}.

The linear behavior of the internal emission yield induced by the potential energy of the projectile is in agreement with previous measurements of the external emission of electrons induced by slow ions of not too high charge states \cite{delauny1987,kurz1993}. For slow ions, deexcitation of the hollow atom and the emission of electrons happens mostly already in front of the surface. If we assume an isotropic emission, half of these electrons are emitted into the solid where they may overcome the tunneling barrier. The mean deexcitation time within the solid is about 10~fs~\cite{Hattass1999,schenkel97}. Thus, the potential energy surviving the pre-surface deexcitation processes will be dissipated within the first few atomic layers. Any excitation of electrons so close to the surface will also give rise to considerable external emission. Therefore, a comparable dependence of the yield on the potential energy is to be expected for external and internal emission. 

\begin{figure}
\includegraphics[width=12cm]{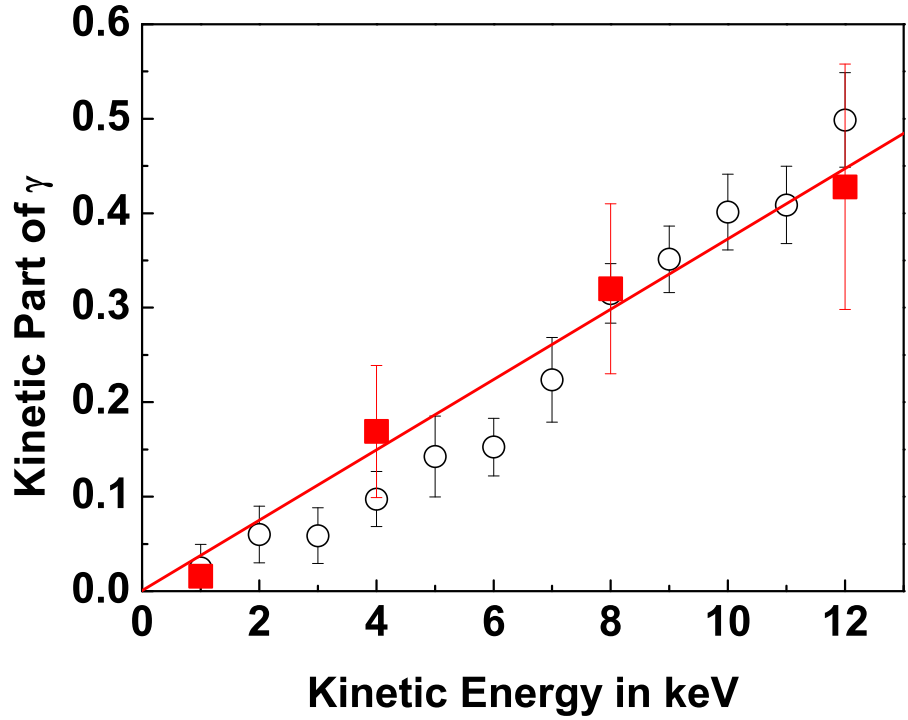}
\caption{Kinetically induced internal emission yield vs. kinetic impact energy of Ar projectiles for different charge states. Open circles represent data for $q=1$ taken from ref.~\cite{meyer04}. Full squares refer to charge states $q>2$ as determined from the extrapolation of data displayed in fig.\ref{alles}.
}
\label{Ekin} 
\end{figure}

The kinetic part of the internal electron emission yield can be determined by extrapolation of the straight lines drawn in fig.~\ref{alles} towards $E_{\mathrm {pot}}= 0$. The resulting values are displayed in fig.~\ref{Ekin} for the four different impact energies investigated here. Due to the fact that the four linear regression lines are parallel, the kinetic yield appears to be independent of the actual projectile charge state $q$. This can be compared to the yield measured for singly charged ($q = 1$) projectiles \cite{meyer04}, which are also included in fig.~\ref{Ekin}. Note, that the data of ref.~\cite{meyer04} were measured for a different impact angle and therefore had to be multiplied by a factor of 2.5~\cite{heuser07} in order to be comparable to the yields measured here. Despite the fact that the data were measured using different MIM junctions, one finds a good overall agreement. 

From fig.~\ref{Ekin} it is seen that the kinetically induced tunneling yield exhibits an approximately linear dependence on the kinetic energy for E$_{\mathrm{kin}} \leq $ 12 keV. From the straight line one finds a yield of about $0.37 \cdot 10^{-4}$ per eV of kinetic impact energy. Note, that this value is by more than an order of magnitude smaller than the potential yield ($\approx 1 \cdot 10^{-3}$eV$^{-1}$) determined above.
Apparently, the efficiency of kinetic energy to produce hot carriers capable of overcoming the tunnel barrier is much smaller than that of potential energy introduced into the surface.

In part, this can be rationalized by the fact that -- for the energy range investigated here -- the energy loss of particles after penetrating the surface is dominated by nuclear rather than electronic stopping. However, target atoms set in motion by elastic collisions also experience inelastic energy loss to the electronic system. In fact, simulations show that more than half of the initial kinetic energy is transiently transferred to the electronic system within the first few ps after the projectile impact \cite{duvenbeck07}. The vast majority of this energy, on the other hand, is stored in low energy excitations, which do not contribute to the tunneling yield \cite{lindenblatt06}. This situation is different for the potential energy carried by the MCI which predominantly generates high energy excitations via Auger deexcitation processes. Hence, hot charge carriers produced by the dissipation of potential energy will be more likely to overcome the tunnel barrier.

Since the tunneling yield is dominated by charge carriers in states located above the barrier one can estimate the fraction $f_t$ of the potential energy $E_{Pot}$ which is carried by the tunneling current as 
$$
f_t\geq\frac{\gamma \cdot \Phi }{E_{\mathrm {pot}}}
$$
where $\Phi $ is the mean height of the tunnel barrier. 
With $\gamma/E_{Pot}\approx 1 \cdot 10^{-3}$eV$^{-1}$ and $\Phi \approx$ 3~eV (see above) this results in $f_t\geq$0.3\% regardless of the charge state $q$.
This value can be compared to the energy fraction released in form of external electron emission (3\% for $q=$ 2 to 15\% for $q=$ 6 \cite{kost07}). The ratio between both fractions can be interpreted as the effective transmission of the MIM top metal layer, which apparently varies between 10\% for $q=$ 2 and 2\% for $q=$ 6. Translated to an effective mean free path, these values correspond to $\lambda$=11~nm and $\lambda$=6.4~nm, respectively. The difference can be understood if we take into account that electrons which are excited by projectiles with lower charge states will on average have lower energy than electrons excited by ions with higher charge states.


In conclusion, the results presented here demonstrate the applicability of MIM junctions as a novel tool to unravel the energy dissipation paths following the impact of multiply charged ions onto a metallic surface. Compared with external electron emission, the MIM provides complementary information on excitation states located below the vacuum level.

The results show quite similar trends, indicating that the tunneling yield is predominantly caused by an over the barrier flux of hot internal charge carriers. Comparison of internal and external energy fluxes reveals an effective mean free path of the generated hot carriers  which is in good agreement with data extracted from a variation of the top silver layer thickness \cite{meyer}.

Furthermore, the tunnel barrier represents an energy dispersive element that can in principle be varied, e.g.~by changing the oxide material or by applying a bias voltage across the junction. Hence, MIM junctions not only provide valuable information on yields of internal hot carriers excited by MCI impact, but can also be used to obtain further information on their spectral distribution. Experiments exploiting these possibilities are currently underway.

\section*{Acknowledgement}
Financial support by the DFG - SFB616: {\it Energy dissipation at
surfaces} and by the European ITS LEIF network RII3\#026015 is gratefully acknowledged.

\clearpage

\end{document}